\documentclass[11pt]{article}
\pdfoutput=1
\usepackage{amsmath}
\usepackage{amssymb}
\usepackage{slashed}
\usepackage{graphicx}
\usepackage{graphics}
\usepackage{epsfig}
\usepackage{soul}
\usepackage{hyperref}
\usepackage{upgreek}
\usepackage{authblk}
\usepackage{blindtext}
\usepackage[section]{placeins}
\usepackage{mathrsfs}
\usepackage{dsfont}

\usepackage{graphicx}

\usepackage[normalem]{ulem}

\usepackage{dcolumn}
\usepackage{bm}

\usepackage{color}
\usepackage[usenames,dvipsnames,svgnames,table]{xcolor}
\usepackage{slashed}
\usepackage{empheq}
\newcommand{\bea}{\begin{eqnarray}}
\newcommand{\eea}{\end{eqnarray}}
\newcommand{\be}{\begin{equation}}
\newcommand{\ee}{\end{equation}}

\usepackage[latin1]{inputenc} 
\title{Editorial: Coarse graining in quantum gravity -- Bridging the gap between microscopic models and spacetime physics}

\author[1]{Astrid Eichhorn \thanks{eichhorn@cp3.sdu.dk}}
\author[2]{Benjamin Bahr \thanks{benjamin.bahr@desy.de}}
\author[3,4]{Antonio D.~Pereira\thanks{a.duarte@science.ru.nl and adpjunior@id.uff.br}}
\affil[1]{CP3-Origins, University of Southern Denmark, Campusvej 55, DK-5230 Odense M, Denmark}   
\affil[2]{II Institute for Theoretical Physics, University of Hamburg, Luruper Chaussee 149, 22761 Hamburg, Germany}
\affil[3]{Institute for Mathematics, Astrophysics and Particle Physics (IMAPP),
Radboud University, Heyendaalseweg 135, 6525 AJ Nijmegen, The Netherlands}
\affil[4]{Instituto de F\'isica, Universidade Federal Fluminese, Av. Litor\^anea s/n, 24210-346, Niter\'oi, RJ, Brazil}

\begin{document}  

\maketitle

\begin{abstract} 
The Renormalization Group encodes three concepts that could be key to accelerate progress in quantum gravity. First, it provides a micro-macro connection that could connect microscopic spacetime physics to phenomenology at observationally accessible scales. Second, it enables a search for universality  classes that could link diverse quantum-gravity approaches and allow us to discover that distinct approaches could encode the same physics in mathematically distinct structures. Third, it enables the emergence of symmetries at fixed points of the Renormalization Group flow, providing a way for spacetime symmetries to emerge from settings in which these are broken at intermediate steps of the construction. These three concepts make the Renormalization Group an attractive method and conceptual underpinning of quantum gravity. Yet, in its traditional setup as a local coarse-graining, it could appear at odds with concepts like background independence that are expected of quantum gravity. Within the last years, several approaches to quantum gravity have found ways how these seeming contradictions could be reconciled and the power of the Renormalization Group approach unleashed in quantum gravity. This special issue brings together research papers and reviews from a broad range of quantum gravity approaches, providing a partial snapshot of this evolving field.
\end{abstract}

\section{Renormalization in quantum gravity -- coming full circle?}
Historically, a large number of approaches to quantum gravity take their starting point in the observation that perturbative renormalizability breaks down for the Einstein Hilbert action \cite{Goroff:1985sz}.  This has triggered developments of various paradigms and frameworks that in a more or less radical sense depart from more standard quantum field theoretic ideas. For instance,  canonical quantization in the presence of a minimal length has been developed in Loop Quantum Gravity \cite{Rovelli:2004tv,Thiemann:2007zz}.  Nonlocality and a corresponding departure from local quantum field theory is implemented in string theory \cite{Polchinski:1998rr,Polchinski:1998rq}. The postulate of Lorentzian discreteness is worked out in causal sets \cite{Surya_2019} and Causal Dynamical Triangulations \cite{Loll_2019}. More examples for various departures from the local quantum field theory framework  -- which is tremendously successful for the other fundamental forces -- can be found. \\
Yet, early on, several ideas already existed on how to formulate quantum gravity as a local quantum field theory with a focus on its Renormalization Group (RG) behavior. These include a translation of a concept from statistical physics to quantum gravity: In the former, interacting RG fixed points encode universality classes, where universal critical behavior emerges from distinct microscopic settings and the physics is determined in terms of a small number of free parameters.
 In \cite{weinberg1979}, Weinberg recognized that an interacting fixed point could encode the ultraviolet (UV) behavior of a quantum field theory. In a similar mechanism to that in statistical physics, asymptotic safety ensures that the physics is encoded in a finite set of free parameters. Moreover, quantum field theoretical ideas for gravity were developed in Stelle's higher derivative gravity \cite{Stelle:1976gc}, which is asymptotically free \cite{Avramidi:1985ki}, but where unitarity is under investigation \cite{Anselmi:2017ygm,Donoghue:2019fcb}. In both settings, (quantum) scale symmetry, emerging at RG  fixed points, is the key to a predictive quantum field theory.  Moreover, asymptotically free or safe theories are well-defined at arbitrarily short length scales and thus constitute candidates for fundamental (in contrast to effective) quantum field theories.
   Yet, the technical tools to comprehensively explore such quantum field theoretic paradigms for quantum gravity still needed further development. In part for this technical reason, alternative paradigms for quantum gravity, e.g., those listed above, were developed. Simultaneously, the concepts tied to the RG, most importantly universality and (quantum) scale symmetry did not play an important role in thinking about quantum gravity.\\

Today, the RG is experiencing a renaissance in quantum gravity.
 This goes significantly beyond the concept of renormalization as a systematic procedure for removing divergences. Rather, the RG plays a central role in the very definition of the quantum theory.
 In short, technical breakthroughs in various formulations of coarse graining enable a search for universality and scale symmetry in a broad range of setups. In the modern formulation of RG tools for quantum gravity, the notion of scale is implemented in several distinct, intricate ways. 
For instance in matrix and tensor models as well as group field theories, the notion of scale is tied to the number of degrees of freedom in a pre-geometric setting. More specifically, the scale is linked to the number of entries of a matrix or tensor in the one case \cite{Brezin:1992yc,Rivasseau:2011hm,Eichhorn:2013isa} or, based on these developments, Fourier-like components on a group manifold \cite{Benedetti:2014qsa} in the other. 
 Besides being manifestly background independent, such ideas can even provide a natural setting to encode a Lorentzian coarse-graining \cite{Eichhorn:2017bwe}. 
As another example, within the holographic  RG, the notion of scale refers to the integrating out of bulk degrees of freedom \cite{Heemskerk:2010hk}.
Both in Loop Quantum Gravity and in the Spin Foam approach, the RG scale refers to the size and complexity of the discrete lattice on which the theory is defined. Here it is the notion of cylindrical consistency of coarse and fine lattices which is used to define an RG flow, at the limit of which sits the continuum theory \cite{Dittrich:2012jq, Bahr:2014qza}.

Finally, background dependent coarse graining on all possible backgrounds simultaneously implements a local and yet background independent notion of scale \cite{Reuter:1996cp,Reuter:2019byg}. \\

 In summary, a range of distinct approaches to quantum gravity are converging towards the point of view that coarse-graining and the associated notion of scale symmetry could enable us to probe properties of quantum space-time. This convergence in itself could prove to be a catalyst for breakthroughs: while every single approach to quantum gravity is facing open questions and challenges, both of conceptual and of technical nature, many insights obtained within the distinct approaches are in fact complementary.  Questions that are seen as technically and/or conceptually challenging in one given approach, might be more easily tackled in another one. Therefore the development of a common language like the RG and the associated unified conceptual framework holds the promise that important insights could be translated between approaches. 

Moreover, there is the distinct possibility that what we now perceive of as different approaches to quantum gravity are in fact simply mathematically different formulations of the same physics. More specifically, different approaches could give rise to the same universality class, thus resulting in the same infrared physics. 

At the same time, a continuum limit encoded in that universality class could ensure the restoration of diffeomorphism symmetry in discrete approaches.
This can, for instance be observed in the case of Regge Calculus and the renormalization of spin foam models \cite{Bahr:2009ku, Dittrich:2014ala, Bahr:2016hwc}.\\

\emph{Future perspectives}: We consider the confluence of quantum-gravity approaches at this RG-vantage point a promising research area that is still in its relatively early stages. 
Promising ideas that have partially been substantiated in proof-of-principle papers within simplified settings include, e.g.,
\begin{enumerate}
\item[i)] the connection to phenomenology, that relies explicitly on the micro- to macro-connection encoded in the RG point of view, see, e.g., \cite{Eichhorn:2018whv,Reichert:2019car,Eichhorn:2020sbo,Gies:2021upb} and references therein, 
\item[ii)] the relation between approaches as members of the same universality class could enable an agreement on physical predictions arising from mathematically distinct frameworks, see, e.g., \cite{Biemans:2016rvp,Eichhorn:2019hsa,Kelly:2019rpx,Benedetti:2020iyz,deBrito:2020rwu,Ambjorn:2021fkp} and references therein,
\item[iii)] the restoration of symmetries that are at the heart of the modern understanding of gravity, namely diffeomorphism symmetry and local Lorentz symmetry that could be recovered in the continuum limit, see, e.g., \cite{Knorr:2018fdu,Mitchell:2020fjy,Asante:2020qpa,Dai:2021fqb} and references therein, even though some approaches break these symmetries at intermediate stages of the construction.
\end{enumerate}
Presently, all three points are at the level of perspectives for the future that are substantiated to varying degrees. They have the potential to transform research in quantum gravity, which motivates a concerted effort in this direction.

The special issue ``Coarse graining in quantum gravity: Bridging the gap between microscopic models and spacetime physics'' provides an incomplete snap-shot of this evolving field, highlighting novel ideas, pointing out open challenges and reviewing recent developments. The diverse perspectives brought together in this issue highlight the broad set of research lines converging towards each other as well as the broad range of research opportunities that are opening up.

\section{List of papers in the special issue}
A series of papers focuses on various aspects of the asymptotic-safety approach, both with continuum and lattice techniques.

 In \cite{Donoghue:2019clr}, J.~Donoghue provides a constructive criticism of the asymptotic-safety program and discusses a key open question of the current state of the art of the asymptotic safety program, namely the Lorentzian signature of space time.
In \cite{Bonanno:2020bil}, the authors critically reflect on the state of the art in asymptotically safe quantum gravity, providing a comprehensive list of open questions and critically reviewing potential pathways to finding answers both within the functional RG approach and lattice techniques. These two papers exemplify the usefulness of constructive criticism across research lines in quantum gravity.

In \cite{Ambjorn:2020rcn}, J.~Ambjorn, J.~Gizbert-Studnicki, A.~Gorlich, J.~Jurkiewicz and R.~Loll use the Causal Dynamical Triangulation approach as a concrete framework to search for asymptotic safety in quantum gravity. They use measurements of the correlation function of the spatial volume profile of the emergent effective spacetime to define lines of constant physics and search for a UV fixed point.

Two papers deal with two aspects of background independence in functional RG techniques for quantum gravity.
In \cite{Pawlowski:2020qer}, J.~Pawlowski and M.~Reichert provide a comprehensive review of fluctuation field calculations in the functional RG framework for asymptotically  safe quantum gravity. The fluctuation field arises within the background field methods that enables the definition of a local coarse graining procedure in a background independent manner. In \cite{Pagani:2020say}, C.~Pagani and M.~Reuter explore the peculiarities and physical implications of background-independent RG flows using field theoretic degrees of freedom like the metric. The requirement of background independence is encoded in self-consistent backgrounds. Both technically and conceptually this constitutes a departure from standard coarse-graining procedures on a fixed background. Specifically, the authors discuss how the so-called naturalness problem regarding the cosmological constant is profoundly different from this perspective.

In \cite{Platania:2020lqb}, A.~Platania reviews developments that tackle a central question in any approach to quantum gravity, namely its phenomenological viability. More specifically, she explores potential consequences of gravitational anti screening, associated with the UV limit in asymptotically safe quantum gravity, in early-universe cosmology. The current state of the art in the field does not yet allow a robust derivation of effects in cosmology from a fundamental gravity theory, instead one can develop and analyze quantum-gravity inspired models. The tentative nature of the link between a fundamental theory and cosmological observations also makes a quantitative comparison between distinct approaches to quantum gravity challenging.
In the absence of experimentally measurable observables, the comparison of characteristic properties of the quantum geometry between approaches to quantum gravity is a potential pathway to find commonalities or differences.
In this spirit, A.~Kurov and F.~Saueressig link functional RG techniques to the analysis of geometric operators and observables in quantum gravity in  \cite{Kurov:2020csd}.

In \cite{Held:2020kze}, A.~Held discusses the notion of effective asymptotic safety. This can be viewed as an additional step in Weinberg's translation of interacting RG fixed points from statistical physics, as the asymptotically safe fixed point is approached at scales that are infrared scales when viewed from a more fundamental model (while they remain UV scales compared to, e.g., the electroweak scale).
In that scenario, an interacting fixed point is therefore approximately realized over an intermediate regime of scales, focusing trajectories starting from different microphysics onto common predictions for the macrophysics. This qualitative idea -- that has the potential to provide a unification between asymptotically safe gravity and other approaches -- is made quantitively precise with a calculable notion of predictive power worked out in detail in gauge-Yukawa models.

Other approaches to quantum gravity in which RG ideas and techniques play a role are then discussed in the following list of papers.

In \cite{Steinhaus:2020lgb}, S.~Steinhaus provides an updated review on the implementation of coarse-graining techniques in spin foam models. He focuses on modern techniques such as tensor network renormalization. Such techniques could allow to search for and probe the continuum limit of such models. In the closely related group field theories, in \cite{Finocchiaro:2020fhl}, M. Finocchiaro and D. Oriti report on the status of simplicial group field theories and their renormalization. In particular, they present a perturbative computation of corrections to correlation functions and provide a road map for the field.

In \cite{Thiemann:2020cuq}, T.~Thiemann takes the canonical quantum gravity approach as a mathematically well-defined setup to tackle the problem of quantizing gravity. He applies the constructive quantum field theory program and, as such, advocates that renormalization is an essential ingredient to fix some ambiguities in the framework. This constitutes a review of a series of papers written by the author and collaborators, where  Hamiltonian Renormalization is established in the canonical language by means of constructive quantum field theory. One of the papers in that series is also published in the present special issue, see \cite{10.3389/fspas.2020.547550}, by K.~Liegener and T.~Thiemann.

A different solution to the coarse-graining problem in gravity is implemented in the discrete, pre-geometric setting of matrix models by T.~Koslowski and A.~Castro. In \cite{Castro:2020dzt},  they apply discrete functional RG techniques to matrix models which encode a preferred foliation. They discover fixed points which are compatible with known results from causal dynamical triangulations in two dimensions  and thereby establish the applicability of these techniques in a causal setting. This work constitutes a compelling example of how numerical Monte-Carlo simulations can meet the computationally less expensive techniques arising from the functional RG.

In \cite{Steinwachs:2020jkj}, C.~Steinwachs reviews developments that aim at formulating quantum gravity as a perturbatively renormalizable and ghost-free quantum field theory, enabled by a breaking of full diffeomorphism invariance to foliation-preserving diffeomorphisms. \\

In summary, these papers provide a partial snapshot of the state-of-the-art of the RG framework to quantum gravity that brings together previously disconnected approaches and is suitable to tackle both formal as well as phenomenological questions.\\

\emph{Acknowledgements}
The authors thank many individuals for discussions on the Renormalization Group perspective in quantum gravity over the years, including J.~Ambjørn, A.~Ashtekar, D.~Benedetti, J.~Ben Geloun, G.~Brito, S.~Carrozza, B.~Dittrich, J.~Donoghue,  L.~Freidel, H.~Gies, R.~Gurau, A.~Held, B.~Knorr, T.~Koslowski, J.~Laiho, D.~Litim, R.~Loll, D.~Oriti, J.~M.~Pawlowski, R.~Percacci, A.~Platania, M.~Reichert, M.~Reuter, V.~Rivasseau, F.~Saueressig, M.~Schiffer, L.~Smolin,  S.~Steinhaus, S.~Surya, T.~Thiemann, C.~Wetterich.

\bibliography{References}

\end{document}